\documentclass[aps,prc,floatfix,superscriptaddress,a4paper,
               showpacs,nofootinbib,preprint]{revtex4}

\usepackage{graphicx}

\usepackage{amsmath}
\usepackage{amssymb}

\newcommand{\be}{\begin{equation}}
\newcommand{\ee}{\end{equation}}
\newcommand{\bel}[1]{\be\label{#1}}
\newcommand{\re}[1]{Eq.~(\ref{#1})}

\newcommand{\hsp}{\hspace*{1pt}}
\newcommand{\hspm}{\hspace*{.5pt}}

\newcommand{\ds}{\displaystyle}
\newcommand{\ov}[1]{\overline{#1}}
\usepackage{bm}

\begin{document}

\title{ New scenarios for hard-core interactions \\in a hadron resonance gas}

\author{L.~M. Satarov}
\affiliation{
Frankfurt Institute for Advanced Studies, D-60438 Frankfurt am Main, Germany}
\affiliation{
National Research Center ''Kurchatov Institute'' 123182 Moscow, Russia}

\author{V. Vovchenko}
\affiliation{
Frankfurt Institute for Advanced Studies, D-60438 Frankfurt am Main, Germany}
\affiliation{
Institut f\"ur Theoretische Physik,
Goethe Universit\"at Frankfurt, D-60438 Frankfurt am Main, Germany}
\affiliation{
Department of Physics, Taras Shevchenko National University of Kiev, 03022 Kiev, Ukraine}

\author{P. Alba}
\affiliation{
Frankfurt Institute for Advanced Studies, D-60438 Frankfurt am Main, Germany}

\author{M.~I. Gorenstein}
\affiliation{
Frankfurt Institute for Advanced Studies, D-60438 Frankfurt am Main, Germany}
\affiliation{
Bogolyubov Institute for Theoretical Physics, 03680 Kiev, Ukraine}

\author{H. Stoecker}
\affiliation{
Frankfurt Institute for Advanced Studies,
D-60438 Frankfurt am Main, Germany}
\affiliation{
Institut f\"ur Theoretische Physik,
Goethe Universit\"at Frankfurt, D-60438 Frankfurt am Main, Germany}
\affiliation{
GSI Helmholtzzentrum f\"ur Schwerionenforschung GmbH, D-64291 Darmstadt, Germany}

\begin{abstract}
The equation of state of a baryon-symmetric hadronic matter with hard-sphere
interactions is studied. It is assumed that mesons $M$ are point-like, but
baryons $B$ and antibaryons $\ov{B}$ have the same hard-core radius $r_B$\hspm .
Three possibilities are considered: 1)~the~$BB$ and $B\ov{B}$ interactions are the same;
2)~baryons do not interact with {antibaryons}; 3)~the $B\ov{B}$, $MB$, and $M\ov{B}$
interactions are negligible. By choosing the parameter $r_B=0.3-0.6$~fm, we calculate
the nucleon to pion ratio as a~function of temperature and perform the fit of
hadron yields measured in central Pb+Pb collisions at $\sqrt{s_{NN}}=2.76~\textrm{TeV}$.
New nontrivial effects in the interacting hadron resonance gas at~temperatures $150-200$~MeV are found.
\end{abstract}

\maketitle
%\centerline\today

\section{Introduction}

The equation of state (EoS) and the phase diagram of the strongly interacting matter
are in the focus of high-energy heavy-ion collisions
and in astrophysics. Some information on infinite equilibrium systems has been
obtained from lattice QCD calculations~\cite{Bor14,Baz14}. However,
that approach has not yet been developed for large baryon chemical potentials
and for small temperatures. The EoS in this domain is still rather uncertain.
Phenomenological models of phase transitions in nuclear matter show~\cite{Ris91,Sat09}
that a realistic phase diagram  cannot be obtained without an explicit account
of the hadronic interactions.

Since the beginning of the 1980ies
several thermal models were constructed~\cite{Stoecker-1981,Stoecker-1987,Cle93,Bra96,Bec01}
to describe yields of secondary hadrons in relativistic nuclear collisions.
These model  assume that such particles are emitted from a statistically
equilibrated system with an ideal gas EoS. The~temperature $T$ and baryon chemical
potential $\mu_B$ of the emitting source were found by fitting the observed yields
of stable hadrons. Many experimental data have been reproduced within this approach.
On the other hand, simple estimates show that hadron densities at the chemical
freeze-out stage of the reaction are rather large, therefore,
one can expect significant deviations from the ideal gas picture.

The hard-sphere interaction is one of the most popular approximations
for implementing short-range repulsion in multiparticle systems, both in
molecular and nuclear physics. It is assumed that particles move freely
unless the distance between their centers equals the sum of their hard-core radii.
This approximation was suggested by van der Waals
for describing properties of dense gases and liquids. Similar
'excluded volume' models were applied~\cite{Ris91,Cle86} to study
the effects of short-range interactions in hadronic systems.
Early versions of this model chose the same hard-core
radius~\cite{Yen97,Whe09,Sat09} for all hadronic species. Attempts to introduce
different radii for different kinds of hadrons have
been made in Refs.~{\cite{Gor99,Bug13a,Bug13b,Vov15a,Vov16}. Later on, more refined versions of the
excluded volume approach were developed which agree well with the
viral expansion~\cite{Lan75} for systems with the hard-sphere interaction.
In particular, the Carnahan-Starling (CS) approximation~\cite{Car69} have
been applied in~\cite{Sat15,Anc15}. As demonstrated in Ref.~\cite{Sat15},
superluminal sound velocities appear in the CS approach only at very high energy
densities, where the deconfinement effects become important.

There is another problem disregarded in existing excluded volume calculations.
They implicitly assume that antibaryons $\ov{B}$ interact
with baryons $B$ in the same way as the baryon-baryon pairs. On the other hand,
there are arguments~\cite{Kle02} that the $B\ov{B}$
interactions should be less repulsive than those for $BB$ pairs. Up to now
not much is known about the~$B\ov{B}$ and meson-(anti)baryon short-range interactions.
Below it is assumed that mesons~$M$ are point-like (with vanishing  hard-core) and
all (anti)baryons have the same hard-core radii. Three possible
scenarios are considered: 1)~the~$BB$ and $B\ov{B}$ interactions are the same;
2)~baryons do not interact with antibaryons; 3)~the $B\ov{B}$, $MB$, and $M\ov{B}$
interactions are neglected. As~far as we know, we are the first who takes into account
possible difference of the $BB$ and~$B\ov{B}$ interactions in the~excluded volume approach.

Most calculations in this paper are done for  baryon-symmetric matter with equal
numbers of baryon and antibaryons, i.e. assuming $\mu_B=0$.
Presumably, such matter is formed in nuclear collisions at the LHC energies.
Using the above mentioned models with the full set of known hadrons,
we fit the midrapidity hadron yields observed by the ALICE Collaboration
in central Pb+Pb collisions at $\sqrt{s_{NN}}=2.76~\textrm{TeV}$.
In agreement with Ref.~\cite{Vov16} we show that~$\chi^2/N_{\rm dof}$ values for
all our fits are  much broader functions of temperature as compared to the ideal
gas calculations.

We found strong effects of hadron short-range interactions in the high temperature region.
It is shown that at $\mu_B=0$, the nucleon to pion ratio as a function of temperature
has a~maximum in the interval $T=150-200~\textrm{MeV}$. The position and height of this
maximum are model dependent. In particular, they are rather sensitive to omitting
the $B\ov{B}$ repulsion. We apply the same models to study the temperature dependence
of the pressure. These results are compared to lattice QCD calculations.

This paper is organized as follows. In Sec.~II we introduce different schemes
to study the effects of the hard-core repulsion in a~one-component gas as well as in
a~multi-component system of hadrons. In Sec.~III
we present numerical results for a baryon-symmetric system containing the full set of
known hadrons. Attention is paid to calculating the nucleon to pion ratio. The results for
pressure as a function of temperature are also presented and compared with lattice QCD data.
Sec.~IV presents our fits to the hadron yields measured by the~ALICE Collaboration.
The conclusions and an outlook are presented in Sec.~V. Appendices~A, B and C provide
formulae for calculating thermodynamic functions in the grand canonical variables.

\section{Hard core repulsion in hadron gas}

In the Boltzmann approximation the ideal gas pressure $P^{\rm id}$ in the grand canonical
ensemble (GC\hsp E) can be written as (\mbox{$\hbar=c=1$})
\bel{idgp}
P^{\rm id}(T,\{\mu\})=T\sum_{i} n^{\rm id}_i(T,\mu_i),~
n^{\rm id}_i(T,\mu_i)=\exp\left(\frac{\mu_i}{T}\right)\phi_{\hsp i}(T),~
\phi_{\hsp i}(T)=\frac{\ds g_i\hspm m_i^2\hspm T}{\ds 2\hspm\pi^2}
K_2\left(\frac{\ds m_i}{\ds T}\right),
\ee
where  $T$ is the system temperature,  $\mu_i$ and $n^{\rm id}_i$ denote, respectively, the
chemical potential and the ideal gas number density
of $i$th hadron species, $m_i$ and $g_i$ are, respectively, the $i$th hadron mass and
statistical weight, $K_2(x)$ is the McDonald function. The function
$\phi_{\hsp i}(T)$ denotes the ideal gas density of $i$th hadrons at $\mu_i=0$.
Note that we apply the zero-width approximation to find contributions of hadronic resonances.

In this paper we study the equilibrium system composed of
hadrons with the hard-core repulsion. Different schemes for taking into account these interactions
are considered. First, the one-component gas is studied, and then the multi-component mixtures of
hadron species will be discussed.

To illustrate the physical effects in different model formulations we first consider
a simple system of nucleons $N$, anti\-nucleons~$\ov{N}$, and pions $\pi$
at temperature $T$ and the baryon chemical potential $\mu_B=0$.
This example will be used to study qualitatively the role of short-range repulsive
interactions in the hadronic system which includes simultaneously mesons and
baryon-antibaryon pairs. In the considered case the chemical potentials of all
hadrons vanish, $\mu_i=0$ ($i=N,\ov{N},\pi$), and the densities of antinucleons
and nucleons coincide, $n_{\ov{N}}=n_N$. In these calculations we assume that nucleons have
a~finite hard-core radius $r_N$ and pions are point-like, i.e. \mbox{$r_\pi=0$}.

\subsection{One-component gas of hard spheres\label{hcr1}}

In this subsection we consider the EoS of a single-component system with the hard-sphere
interaction of particles. Let us denote by $r$ the hard-core radius of the particle.
In the Boltzmann approximation, one can write the following 'exact' expression for
the pressure~\cite{Mul08}:
\bel{pbm}
P=n\hspm T\hspm Z\hspm (\eta)\,.
\ee
Here $n$ is the number density of particles, $\eta=n\hspm v$ denotes their ''packing''
fraction, where $v=4\pi r^3/3$ is a~single-particle hard-core eigenvolume.
The dimensionless ''compressibility'' factor~$Z\hspm (\eta\hspm )$ does not depend on
temperature. It is clear that $Z\hspm (\eta\hspm )\to 1$ in
the ideal gas limit~$\eta\to 0$\hspm . Note that~\re{pbm} is valid for packing
fractions below the critical value of the liquid-solid
transition $\eta_{\hspm c}\simeq 0.49$~\cite{Mul08}.

Different authors either use numerically tabulated values of $Z(\eta)$ or apply
analytical approximations. The simple van der Waals excluded volume approximation,
\bel{zem}
Z_{\hsp\rm EV}(\eta)=(1-4\hspm\eta)^{-1}=1+4\hspm\eta+16\hsp\eta^2+\ldots ~,
\ee
is used in the eigenvolume (EV) models. The approximation (\ref{zem}) correctly
describes the second  term of the virial expansion for the pressure~\cite{Lan75}, but it
fails to reproduce higher-order terms which give the contribution of non-binary interactions.
A comparison with numerical calculations shows
that~\re{zem} strongly overestimates the values of~$Z\hsp (\eta)$
at~$\eta\gtrsim 0.2$~\cite{Sat15}. Note that values $\eta>0.25$ are not allowed
in this model.

A very accurate and relatively simple approximation
was suggested~\cite{Car69} by~Carnahan and Starling (CS). It has the form
\bel{zcs}
Z_{\hsp\rm CS}\hspm (\eta)=\frac{\ds 1+\eta+\eta^{\hspm 2}-
\eta^{\hsp 3}}{\ds (1-\hspm\eta)^{\hsp 3}}=1+4\hspm\eta+10\hsp\eta^{\hspm 2}+\,\ldots~.
\ee
Note, that the third term of the virial expansion for $Z\hspm (\eta)$
is correctly reproduced within the~CS model. It is interesting that Eq.~(\ref{zcs})
reproduces rather accurately the virial expansion terms up to the eighth order~\cite{Mul08}.
In fact, this approximation can be safely used in the whole domain $\eta<\eta_{\hspm c}$\hspm .
One can see that $Z_{\hsp\rm CS}\hspm (\eta)\simeq Z_{\hsp\rm EV}\hspm (\eta)$ at small~$\eta$.

The above equations correspond  to the canonical ensemble (CE).
The transformation to the GC\hsp E can be done using the
following procedure. Integrating Eq.~(\ref{pbm}) over the system volume one obtains
the free energy density $f=f(T,n)$ (see Appendix~A). Using the thermodynamical relation
$\mu=~\left(\partial f/\partial n\right)_T$,
one finds the transcendental equation for the GC\hsp E particle density $n=n\hspm (T,\mu)$
\bel{ngce}
n=n^{\rm id}\big[T,\mu-T\psi\hspm (v\hspm n)\hspm\big]\,,
\ee
where $n^{\rm id} (T,\mu)$ is given by the second equality of (\ref{idgp}) with
the replacement $\mu_i\to\mu$. The dimensionless function
\bel{psi}
\psi\hspm (\eta)= Z\hspm (\eta)-1+\int\limits_0^{\eta}\frac{d\eta^{\,\prime}}
{\eta^{\,\prime}}\,\big[Z\hspm (\eta^{\,\prime})-1\hsp\big]
\ee
describes the shift of chemical potential (in units of $T$)
for a one-component matter with hard-sphere interactions as compared to the ideal gas~\cite{Mul99}.
Finally, the~GC\hspm E pressure is calculated by substituting the solution of (\ref{ngce})
into \re{pbm}.

Within the EV and CS models one can calculate the function
$\psi\hspm (\eta)$ analytically. Substituting~(\ref{zem}) and~(\ref{zcs}) into~\re{psi}
gives\hspm\footnote
{
Note that $\psi_{\hsp\rm CS}(\eta)<\psi_{\hsp\rm EV}(\eta)$ at $\eta<0.25$\hsp .
}
\begin{eqnarray}
&&\psi_{\hsp\rm EV}(\eta)=\frac{4\hspm\eta}{1-4\hspm\eta}-\ln{(1-4\hspm\eta)}=
8\hspm\eta+24\hsp\eta^{\hspm 2}+\,\ldots~,\label{psiem}\\
&&\psi_{\hsp\rm CS}(\eta)=\frac{3-\eta}{(1-\eta)^{\hspm 3}}-3=
8\hspm\eta+15\hsp\eta^{\hspm 2}+\ldots~.\label{psics}
\end{eqnarray}
An equivalent GC\hspm E formulation of the EV model was obtained
earlier in~Ref.~\cite{Ris91}.

\subsection{Diagonal eigenvolume model\label{sdem}}
A simple extension of the EV model for multi-component systems was suggested
in~Ref.~\cite{Yen97}.
The pressure of a hadronic mixture is parameterized as
\bel{pds}
P=T\sum\limits_i \xi_i,~~~~~~\xi_i=\frac{n_i}{1-\sum_j b_j\hspm n_j}\,,
\ee
where $b_i=16\pi r_i^3/3$ and $n_i$ are, respectively,
the eigenvolume parameter and the density of~$i$th hadrons.
The sums in~\re{pds} go over all types of hadrons.
Following Ref.~\cite{Vov16}, we denote this excluded volume scheme as the
''diagonal'' eigenvolume model (DEM).

Calculating the free energy density of the hadronic mixture
and taking its derivatives with respect to $n_i$ (see Appendix~A), one finds the
following equation for the chemical potential of $i$th particles:
\bel{cpdem}
\mu_{\hspm i}=T\ln{\frac{\xi_i}{\phi_i\hspm (T)}}+b_i\hsp P\,,
\ee
where $\xi_i$ and $P$ are taken from~\re{pds}. In fact,
the above equation provides the transition from the CE to the~GC\hspm E.
Using~\re{idgp} one can rewrite (\ref{cpdem}) in the equivalent form
\bel{xids}
\xi_i=n^{\rm id}_i\hspm (T,\mu_{\hspm i}-b_i\hspm P)\,.
\ee

Substituting (\ref{xids}) into~\re{pds} gives the transcendental equation
for the pressure~\cite{Yen97}
\bel{pdevm}
P=\sum_i ~P_i^{\rm id}\left(T,\mu_{\hspm i}-b_i\hspm P\right),
\ee
and allows to calculate the particle number densities
\bel{eqni}
n_i=\frac{\xi_i}{1+\sum_j b_j\,\xi_j}\,.
\ee
Equations (\ref{xids}) and (\ref{eqni}) lead to the following expressions for
density ratios of different hadronic species:
\bel{hdr}
\frac{\ds n_i}{\ds n_j}=\frac{n_i^{\rm id}}{n_j^{\rm id}}\,
\exp\left[(b_j-b_{\hsp i})\,\frac{P}{T}\right]\,.
\ee
The  ratio (\ref{hdr})  is smaller than that of the ideal gas if $b_i>b_j\,$.

For the $N\ov{N}\pi$ mixture with $\mu_N=\mu_{\ov{N}}=\mu_{\hspm\ds\pi}=0$ and
''point-like'' pions ($r_{\ds\pi}=0$) one can represent \re{pdevm} in the form
\bel{pds2}
P=T\left[2\hspm\phi_N\hspm (T)\hspm\exp{\left(-\frac{b_N\hspm P}{T}\right)}+
\phi_{\hsp\ds\pi}\hspm (T)\hsp\right].
\ee
Note that the second term on the right hand side (r.h.s.) of~this equation
gives the partial pressure of pions which is not suppressed
as compared to the ideal pion gas.
Equation~(\ref{eqni}) yields the following~expressions for the hadronic densities:
\bel{hdr1}
n_N=\frac{\xi_N}{1+2\hspm b_N\hspm\xi_N}\,,~~~n_{\ds\pi}=\frac{\phi_{\hsp\ds\pi}\hspm (T)}
{1+2\hspm b_N\hspm\xi_N}\,,
\ee
where $\xi_N=\phi_N\hspm (T)\,\exp{(-b_N P/T)}$.
Finally, we arrive at the equation for~the nucleon-to-pion ratio
\bel{npir}
{\frac{N}{\pi}\equiv
\frac{\ds n_N}{\ds n_{\ds\pi}}=\frac{\phi_N\hspm(T)}{\phi_{\hsp\ds\pi}\hspm (T)}\,\,
\exp\left(-~\frac{b_{\hsp N}P}{T}\right)},
\ee
where $P$ is determined by solving~\re{pds2}. Note that (\ref{npir})
is a particular case of~\re{hdr}.

\begin{figure*}[bht!]
\centerline{\includegraphics[trim=0 7.5cm 0 8.5cm, clip, width=0.65\textwidth]{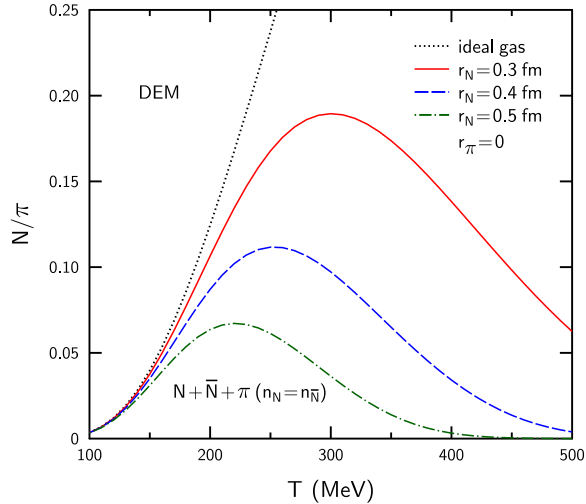}}
\caption[]{(Color online)
The $N/\pi$ ratio as a function of temperature in the $N\ov{N}\pi$ matter with
point-like pions and equal numbers
of nucleons and antinucleons. The solid, dashed, and dash-dotted lines correspond to
nucleon hard-core radii $r_N=0.3, 0.4$ and $0.5~\textrm{fm}$, respectively. The dotted line
is obtained in the ideal gas limiting case $r_N=0$.
}\label{f1:nrpi-ds}
\end{figure*}
Figure~\ref{f1:nrpi-ds} shows {$N/\pi$} ratio in the DEM for different
values of $r_N$ from 0.3 to 0.5 fm. This ratio is a non-monotonic function of $T$
with a maximum in~the temperature range between about 200 to 300 MeV.
Note the same $N/\pi$ ratio is reproduced by two different values
of~$T$. The higher temperature value corresponds to a denser state of the $N\ov{N}\pi$ matter
with stronger short-range interactions of hadrons. It should be noted
that this simple $N\ov{N}\pi$ system is considered here
for illustration. Hadronic states of higher masses will be introduced below.
One should also have in mind the appearance of the crossover transition to the
deconfined quark-gluon plasma at high temperatures.

At temperatures $m_{\ds\pi}\lesssim T\ll m_N$ the
following approximate relations hold (see~\re{idgp}):
\bel{didg}
\phi_N\hspm (T)\simeq\frac{1}{2}\left(\frac{2\, m_NT}{\pi}\right)^{3/2}\,
\exp\left(-~\frac{m_N}{T}\right)\,,~~~~~\phi_{\hsp\ds\pi}\hspm (T)\simeq\frac{3}{\pi^2}\hsp T^{\hsp 3},
\ee
Substituting (\ref{didg}) into (\ref{npir}) gives
\bel{npir1}
\frac{\ds n_N}{\ds n_{\ds\pi}}\simeq\frac{\sqrt{2\pi}}{3}\left(\frac{m_N}{T}\right)^{3/2}
\exp{\big[-\varphi\hsp (T)\big]}\,,
\ee
where
\bel{phif}
\varphi\hsp (T)=\frac{m_N+b_N\hspm P}{T}\simeq\frac{m_N}{T}+\frac{3\hsp b_NT^3}{\pi^2}\,.
\ee
The last estimate is obtained by neglecting the first term in~\re{pds2}.
This is a reasonable approximation at the temperatures considered here.
The temperature dependence of $n_N/n_{\ds\pi}$ is determined mainly by the
last factor in~\re{npir1}\hspm . As one can see from~\re{phif}, it is a non-monotonic function
of $T$ with a maximum at
$T\simeq\left(\frac{\ds\pi^2}{\ds 9}\frac{\ds m_N}{\ds b_N}\right)^{1/4}$\hspace*{-2mm}.
The~calculation shows that the maximum is shifted
from about 350 to 250 MeV when $r_N$ increases from 0.3 to~0.5~fm. This agrees
with the numerical results shown in Fig.~\ref{f1:nrpi-ds}. As will be shown in Sec.~\ref{shrg},
the inclusion of heavier hadrons and resonances reduces the~$N/\pi$ ratio and shifts the maxima
to lower temperatures.

\subsection{Non-diagonal eigenvolume model}

The DEM considered in preceding section is not accurate already in the second order of the virial
expansion for classical particles with hard-sphere interactions. Indeed, expanding \re{pds} in powers of
partial densities leads to the relation
\bel{pdsvir}
\frac{\ds P}{\ds T}=\sum\limits_i n_i+\sum\limits_{i,j}b_{j}n_i n_j\,+\ldots\,,
\ee
whereas the virial expansion gives~\cite{Lan75}
\bel{vire}
\frac{\ds P}{\ds T}=\sum\limits_i n_i+\sum\limits_{i,j}B_{ij}n_i n_j\,+\ldots\,,~~~~~~~
B_{\hsp ij}=2\hspm\pi\hspm (r_i+r_j)^3/3\,.
\ee
One can see that \mbox{$B_{\hsp ij}\neq b_j$} for non-equal hard-core radii,
\mbox{$r_i\neq r_j$}\hsp .
\begin{figure*}[htb!]
\centerline{\includegraphics[trim=0 7.5cm 0 9cm, clip, width=0.65\textwidth]{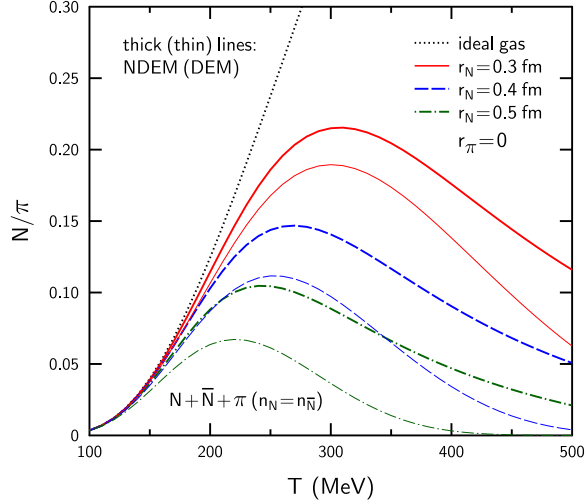}}
\caption[]{(Color online)
The $N/\pi$ ratio as a function of temperature in the $N\ov{N}\pi$ matter with equal numbers
of nucleons and antinucleons. The thick solid, dashed, and dash-dotted lines are calculated
by using the NDEM with nucleon hard-core radii $r_N=0.3, 0.4$ and $0.5~\textrm{fm}$, respectively.
Thin lines give corresponding results obtained within the DEM. The dotted line
is calculated in the ideal gas limit~$r_N=0$.
}\label{f2:nrpi-nds}
\end{figure*}

In this subsection we consider the EoS of
a hadronic mixture by using  the improved ''non-diagonal'' eigenvolume model\hspm\footnote
{
It is called the ''crossterms'' eigenvolume model in Ref.~\cite{Vov16}.
}~(NDEM)  suggested in Ref.~\cite{Gor99}. This scheme yields agreement with the second-order
virial expansion (\ref{vire}).

The pressure in the NDEM is given by~\re{pds} with the replacement $b_j$
by the matrix
\bel{mbij}
b_{\hsp ij}=\frac{2\hspm B_{ij}\hspm B_{ii}}{B_{ii}+B_{jj}}\,.
\ee
Then one gets the relations
\bel{pnds}
P=T\sum\limits_i \xi_i,~~~\xi_i=\frac{n_i}{1-\sum_j b_{\hsp j\hspm i}\hsp n_j}\,,
\ee
Instead of Eqs. (\ref{cpdem}) and (\ref{xids}), after the transition to the CGE
(see Appendix~A) one obtains the equations
\bel{cpnds}
\mu_{\hsp i}=T\left(\ln{\frac{\xi_i}{\phi_i\hspm (T)}}+\sum\limits_j b_{\hsp ij}\,\xi_j\right)\,,
\ee
which are equivalent to
\bel{xinds}
\xi_i=n^{\rm id}_i\hspm (T,\mu_{\hsp i}-T\sum\limits_j b_{\hsp ij}\,\xi_j)\,.
\ee
Note that in a general case one should explicitly solve the set of coupled
equations (\ref{xinds}) instead of a single equation~(\ref{pdevm}) in the DEM.
At known $\xi_i$ one can calculate the densities~$n_i$ by using the second equality in (\ref{pnds}).
The NDEM is reduced to the DEM if all hard-core radii are equal, $r_i=r_j$, and thus
$b_{\hsp ij}=b_{\hsp i}$. In this case, the particle number ratios $n_i/n_j$ become equal to their
ideal gas values $n_i^{\rm id}/n_j^{\rm id}$.

Let us again consider the $N\ov{N}\pi$ system with $\mu_N=\mu_{\ov{N}}=\mu_\pi=0$ and assume
that pions are point-like. Denoting (anti)nucleons and pions by indices '1' and '2', respectively,
one can write the relations $b_{11}=4\hspm b_{12}=b_N, b_{22}=b_{21}=0$\hspm .
As a result, instead of Eqs.~(\ref{pnds}) and~(\ref{xinds}), one
obtains\hspm\footnote
{
One gets the corresponding equations of the DEM after replacing
$\phi_{\hsp\ds\pi}\to 4\hspm\phi_{\hsp\ds\pi}$ in~\re{dtnds2} and
$n_N\to 4\hspm n_N$ in the second equality of~\re{dnds2}.
}
\begin{eqnarray}
&&\frac{P}{T}=2\,\xi_N+\phi_{\hsp\ds\pi},\label{prnds2}\\
&&\xi_N=\phi_{N}\hsp\exp\left[-b_N\hsp \left(2\hsp\xi_N+
\phi_{\hsp\ds\pi}/4\right)\right],\label{dtnds2}\\
&&n_N=\frac{\xi_N}{1+2\hsp b_N\xi_N},~~n_{\ds\pi}=
\phi_{\hsp\ds\pi}\,(1-b_N\hspm n_N/2)\,.\label{dnds2}
\end{eqnarray}
Solving Eq.~(\ref{dtnds2}) with respect to $\xi_N$, we can calculate $P$
and the number densities $n_N, n_{\ds\pi}$ by using Eqs.~(\ref{prnds2})
and (\ref{dnds2}).

A comparison of the $N/\pi$ ratios in the NDEM and DEM is
shown in~Fig.~\ref{f2:nrpi-nds}. It~is seen that at fixed $r_N$ the NDEM predicts larger values
of the $N/\pi$, which are smaller suppressed as compared to the ideal gas.
Below  only the NDEM is used, and it is denoted for brevity the `eigenvolume model' (EVM).

\subsection{Binary mixture of hard spheres and point-like particles\label{scsm}}

In the case of a binary mixture\hsp\footnote
{
One may consider the $N\ov{N}\pi$ matter with $n_N=n_{\ov{N}}$
as a binary $N\pi$ mixture with the 'nucleon' density~$2\hsp n_N$.
However, this is not correct if $NN$ and $N\ov{N}$ interactions are different (see Sec.~\ref{nnpw}).
}
where particles of one component are point-like, it is possible
to apply the CS model (CSM)~\cite{Sat15} which is valid at much larger
densities than the~NDEM.

Let us denote the components of such a binary matter by indices $i=1,2$ and assume
that particles of the first component interact as hard spheres of the radius $r_1$, but
particles of the second kind are point-like. In this case, similar to Eq.~(\ref{pbm}),
one can write the equation~\cite{Mul08}
\bel{pbm1}
P=T\left[n_1\hsp Z\hspm (\eta_1)+\frac{n_2}{1-\eta_1}\right].
\ee
Here $n_i$ is the number density of the $i$\hspm th component, $\eta_1=n_1v_{\hspm 1}$ is the
''packing'' fraction of the first particles where $v_{\hsp 1}=4\pi r_1^3/3$ denotes their single-particle
(hard-core) eigenvolume. The~denominator in the second term \mbox{in the r.h.s. of~\re{pbm}} describes the
reduction of the~volume available for particles $i=2$.

As shown in Appendix~B,~\re{pbm1} leads to following equations for chemical potentials:
\begin{eqnarray}
&&\mu_1=T\left[\ln{\frac{n_1}{\phi_1(T)}}+\psi(n_1v_{\hspm 1})+
\frac{n_2\hspm v_{\hspm 1}}{1-n_1v_{\hspm 1}}\right],\label{cpot1}\\
&&\mu_2=T\left[\ln{\frac{n_2}{\phi_2(T)}}-\ln{(1-n_1v_{\hspm 1})}\right],\label{cpot2}
\end{eqnarray}
where $\psi(\eta)$ is defined in~\re{psi}. The first term in the r.h.s.~of~(\ref{cpot1})
equals the ideal gas chemical potential $\mu^{\hspm\rm id}_1$. The second one
gives the shift of $\mu_1$ induced by interactions of the first particles. The last term
appears due to interactions of particles $i=1$ and $i=2$\hspm . It equals the minimal work
for creating a cavity with the volume $v_1$ inside of a gas of point-like
particles~\cite{Sat15}. The shift of $\mu_2$ is due to reducing the total volume accessible
for parti\-cles~$i=2$\hspm . Using Eqs.~(\ref{psi}) and (\ref{pbm1})--(\ref{cpot2}) one can
prove the validity of the thermodynamic relation $dP=n_1\hspm d\mu_1+n_2\hspm d\mu_2$ for
an arbitrary isothermal process\hspm\footnote
{
This relation provides the thermodynamic consistency of the model.
}. Note the the shifts of chemical potentials disappear in the limit $v_{\hspm 1}\to 0$\hspm .

In the case of vanishing chemical potentials, $\mu_1=\mu_2=0$, one gets
the following relations for the particle densities:
\begin{eqnarray}
&&n_1=n_1^{\rm id}\hspm\exp\left[\ds-\psi\hspm (n_1v_{\hspm 1})-
n_{\hsp 2}^{\rm id}\hsp v_{\hspm 1}\hspm\right],\label{ede1}\\
&&n_2=n_{\hsp 2}^{\rm id}\hsp (1-n_1v_{\hspm 1})\,.\label{ede2}
\end{eqnarray}
Therefore, the problem is reduced to solving~\re{ede1} with respect to $n_1$.

\begin{figure*}[htb!]
\centerline{\includegraphics[trim=0 7.5cm 0 9cm, clip, width=0.65\textwidth]{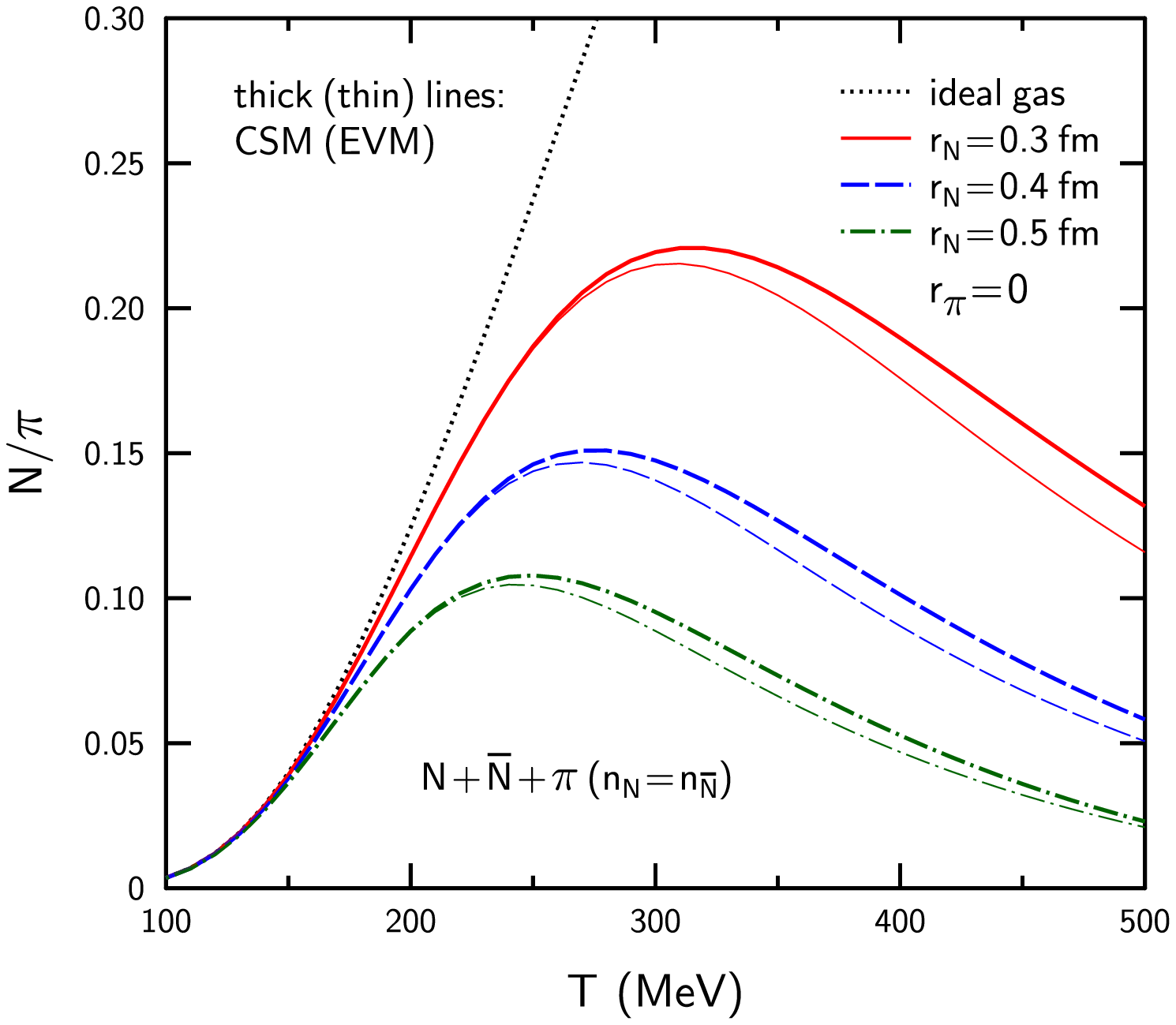}}
\caption[]{(Color online)
The $N/\pi$ ratio as a function of temperature in the $N\ov{N}\pi$ matter with equal numbers
of nucleons and antinucleons. The thick solid, dashed, and dash-dotted lines are calculated
by using the CSM with nucleon hard-core radii $r_N=0.3, 0.4$ and $0.5~\textrm{fm}$, respectively.
Thin lines give corresponding results obtained within the EVM. The dotted line
is calculated assuming~$r_N=0$.
}\label{f3:nrpi-csm}
\end{figure*}
To determine properties of $N\ov{N}\pi$ matter with \mbox{$n_N=n_{\ov{N}}$} one should
substitute \mbox{$n_1=2\hspm n_N$}, \mbox{$n_1^{\rm id}=2\hspm \phi_N$},
\mbox{$v_{\hspm 1}=v_N=b_N/4$}
and make the replacement $n_2\to n_{\hsp\ds\pi}$. One then arrives at the equations
\begin{eqnarray}
&&\frac{P}{T}=2\hspm n_N\hspm Z\hspm (2\hspm n_N\hspm v_N)+\phi_{\hsp\ds\pi},\label{prcss}\\
&&n_N=\phi_N\hsp\exp\left[\ds -\psi\hsp (2\hspm n_N\hspm v_N)-
\phi_{\hsp\ds\pi}\hspm v_N\right],\label{dncss}\\
&&n_{\ds\pi}=\phi_{\hsp\ds\pi}\hsp (1-2\hspm n_N\hspm v_N)\,.\label{dpcss}
\end{eqnarray}
The CSM  is obtained by substituting~the expressions
\mbox{$Z=Z_{\rm CS}, \psi=\psi_{\hspm\rm CS}$} from Eqs.~(\ref{zcs}) and~(\ref{psics}),
whereas the~EVM corresponds to $Z=Z_{\rm EV}, \psi=\psi_{\hspm\rm EV}$
given by Eqs.~(\ref{zem}) and (\ref{psiem}). In~the latter case the above equations
are equivalent to Eqs.~(\ref{prnds2})--(\ref{dnds2}) of the NDEM.

As mentioned above, \mbox{$\psi_{\hspm\rm CS}(\eta)<\psi_{\hspm\rm EM}(\eta)$}.
According to~Eqs.~({\ref{dncss}) and ({\ref{dpcss}), this implies that the
inequalities \mbox{$n_N^{(\rm CSM)}>n_N^{(\rm EVM)}$} and \mbox{$n_{\ds\pi}^{(\rm CSM)}<n_{\ds\pi}^{(\rm EVM)}$}
hold at fixed~$T$ and $r_N$. Therefore, the~$N/\pi$ ratio should be larger in the CSM as compared to the EVM.

Such a conclusion is confirmed by the results of numerical calculations
shown in~Fig.~\ref{f3:nrpi-csm}. However, one can see that the difference
between the CSM and EVM results is not very significant. This follows from
relatively small packing fractions for (anti)nuc\-leons
(\mbox{$\eta=2\hsp n_N\hspm v_N\lesssim 0.1$}) in the $N\ov{N}\pi$ matter at
$\mu_B=0$\hsp . It will be shown below that a similar situation occurs
after inclusion of hadronic resonances. Note, that much larger baryon
densities may be achieved in baryon-asymmetric systems with non\-zero~$\mu_B$~\cite{Sat09}.

\subsection{The $\bm N\ov{\bm N}\bm\pi$ matter without $\bm N\ov{\bm N}$ interactions\label{nnpw}}

Up to now it was assumed that antinucleons interact with nucleons as hard spheres of the same radii,
i.e. we did not introduce any differences between the $N\ov{N}$ and $NN$ interactions.
However, due to the \mbox{G-parity} symmetry, the vector part of the $N\ov{N}$ pair potential
has an opposite sign as compared to the $NN$ one~\cite{Kle02}. As a consequence, the $N\ov{N}$
interaction should be less repulsive at short distances than that for $NN$ pairs\hsp\footnote
{
Motivated by these
features, we introduced in~\cite{Mis05} an attractive vector field for antibaryons
in nuclear matter, predicting strong binding and compression effects for antibaryon-doped nuclei.
}.
Another reason for reduced short-range repulsion is the Pauli exclusion principle which
should be less restrictive for $N\ov{N}$ interactions. Possibility of vanishing short-range
repulsion between baryons and antibaryons has been pointed out in~Ref.~\cite{And12}.

\begin{figure*}[ht!]
\centerline{\includegraphics[trim=0 7.5cm 0 8.5cm, clip, width=0.65\textwidth]{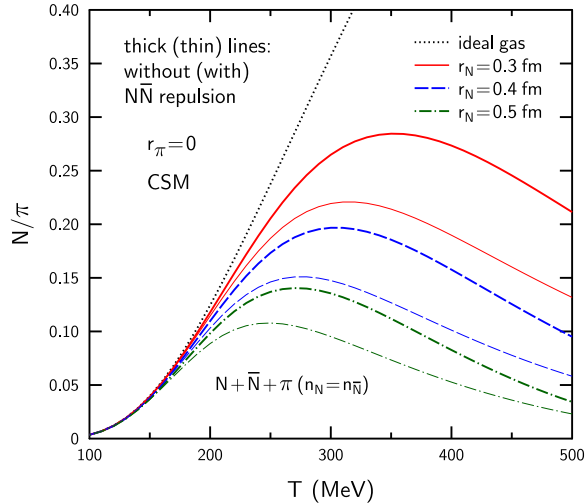}}
\caption[]{(Color online)
The $N/\pi$ ratio as a function of temperature in the $N\ov{N}\pi$ matter with equal numbers
of nucleons and antinucleons. The thick lines are calculated
using the CS approximation of compressibility and disregarding the $N\ov{N}$ interactions.
Thin lines give corresponding results with inclusion of the $N\ov{N}$ repulsion.
The dotted line is calculated in the ideal gas limit $r_N=0$.
}\label{f4:nrpi-nrep}
\end{figure*}
To study the sensitivity to the asymmetry between
 the~$N\ov{N}$ and $NN$ interactions, we consider below the EoS
 of the $N\ov{N}\pi$ matter assuming a totally vanishing repulsion for $N\ov{N}$ pairs.
 In this case the nucleon and antinucleon components of the $N\ov{N}\pi$ system become mutually
 independent. It is not difficult to modify the ''binary mixture'' EoS for the case of $N\ov{N}\pi$
 matter without the $N\ov{N}$ interaction. Instead of the first term in~\re{pbm1} we take the sum of two
 contributions from purely nucleon and antinucleon fluids:
 $$
 (n_N+n_{\ov{N}})\hspm Z(n_Nv_N+n_{\ov{N}}v_N)\to n_N\hspm Z(n_Nv_N)+n_{\ov{N}}\hspm Z(n_{\ov{N}}v_N).
 $$
 Using further the procedure analogous to that used in Sec.~\ref{scsm} one gets the equations which differ
 from Eqs.~(\ref{prcss})--(\ref{dpcss}) by the
 replacement $2\hspm n_N\hspm v_N\to n_N\hspm v_N$ in the arguments of functions~$Z$
 and $\psi$:
\begin{eqnarray}
&&\frac{P}{T}=2\hspm n_N\hspm Z\hspm (n_N\hspm v_N) +\phi_{\hsp\ds\pi},\label{prcss1}\\
&&n_N=\phi_N\hsp\exp\left[\ds -\psi\hspm (n_N\hspm v_N)-
\phi_{\hsp\ds\pi}\hspm v_N\right],\label{dncss1}\\
&&n_{\ds\pi}=\phi_{\hsp\ds\pi}\hsp (1-2\hspm n_N\hspm v_N)\,.\label{dpcss1}
\end{eqnarray}
By comparing Eqs. (\ref{dncss}) and (\ref{dncss1}), one can see that omission
of the $N\ov{N}$ repulsion increases the nucleon density at fixed~$T$ and $r_N$. This corresponds
to effective reduction of the parameter~$r_N$ as compared to standard calculations with
equal $NN$ and $N\ov{N}$ interactions.

The results of both calculations are compared in Fig.~\ref{f4:nrpi-nrep}
where the CS approximation of the compressibility factor is used.
Note that omitting the $N\ov{N}$ repulsion, indeed, leads
to a~significant increase of the $N/\pi$ ratio for large $T$.
Below we extend this analysis to hadronic matter containing heavier hadrons.
In particular, we study the sensitivity of the EoS and particle ratios to the
omission of the short range repulsion between baryons and antibaryons.

\section{Hadron resonance gas\label{shrg}}

As already mentioned, the pure $N\ov{N}\pi$ matter is unrealistic at large
temperatures and one should take into account excitation of
heavier hadrons and hadronic resonances. Note that the presence of resonances
in the hadron gas effectively takes into account the attractive interactions
of hadrons~\cite{Ven92}.	

In this section we consider the EoS of the $B\ov{B}M$ matter which contains
the full set of known baryons ($B$), antibaryons ($\ov{B}$) and mesons ($M$)
with masses below~2.6~GeV. Speci\-fically, we use the set of uncharmed hadrons
included in the THERMUS model~\cite{Whe09}. As~before, we neglect the isospin
effects and deviations from the Boltzmann statistics. Unless stated otherwise,
we apply the zero-width approximation
for all hadronic resonances.

One can easily generalize the approach developed in preceding section,
assuming that all (anti)baryons have the same hard-core radii,
$r_i=r_B~(i\in B,\ov{B})$, and all mesons are point-like, $r_i=0~(i\in M)$.
Below we calculate the total (final) densities of stable hadrons~($\ov{n}_i$)
by including the feeding from strong decays of resonances:
\bel{prdn1}
\ov{n}_i=n_i+\sum\limits_{\bm j}n_j\,B\hspm r\hspm (j\to i)\,,
\ee
where the sum is taken over all resonances. The quantity
$B\hspm r\hspm (j\to i)$ denotes the average number of $i$\hspm th hadrons from
the decays of $j$\hspm th resonance. We calculate these numbers by using the decay
tables given in the THERMUS model.

\subsection{Hadronic matter with baryon-antibaryon repulsion\label{wbbr}}

Let us first assume that there is no difference between the $BB$ and $B\ov{B}$
short-range interactions (the corresponding approach will be referred as the CI calculation).
Below we denote by $n_B, n_{\ov{B}}$ and $n_M$ the total densities of baryons, antibaryons
and mesons, respectively. Applying~\re{pbm1} for the full set of hadrons, one has
\bel{pec1}
\frac{P}{T}=n_{\hspm T}\hspm Z\hspm (n_{\hspm T}\hspm v)+
\frac{n_M}{1-n_{\hspm T}\hspm v}\,,
\ee
where \mbox{$n_T=n_B+n_{\ov{B}}$}, \mbox{$v=4\pi r_B^3/3$} is the hard-core eigenvolume
of a baryon, and~\mbox{$Z=Z\hspm (\eta)$} is the compressibility factor introduced in Sec.~\ref{hcr1}.

As shown in Appendix~C, \re{pec1} leads to the following expressions for chemical
potentials of (anti)baryons and mesons:
\begin{eqnarray}
&&\frac{\mu_i}{T}=\ln{\frac{n_i}{n_i^{\rm id}}}+\psi\hspm (n_{\hspm T}\hspm v)+
\frac{n_M\hspm v}{1-n_{\hspm T}\hspm v}\,,~~~i\hspm\in B,\ov{B},\label{mub1}\\
&&\frac{\mu_i}{T}=\ln{\frac{n_i}{n_i^{\rm id}}}-\ln{(1-n_{\hspm T}\hspm v)}\,,
\hspace*{1.9cm}i\hspm\in M,\label{mum1}
\end{eqnarray}
where $n_i^{\rm id}$ and $\psi(\eta)$ are defined in Eqs.~(\ref{idgp}) and (\ref{psi}),
respectively. Note that first terms in these equations give the chemical potentials of
$i$th hadrons in the ideal gas limit $v\to 0$\hspm .

For the chemically equilibrated matter with zero net baryon charge one has $\mu_i=0$ for
all hadronic species. In this case the relations $n_B=n_{\ov{B}}=n_{\hspm T}/2$ hold.
Using further \mbox{Eqs.~(\ref{pec1})--(\ref{mum1})} one gets the equations
(cf.~(\ref{prcss})--(\ref{dpcss}))
\begin{eqnarray}
&&\frac{P}{T}=2\hspm n_B\hspm Z\hspm (2\hspm n_{\hspm B}\hspm v)+
n_M^{\rm id}\,,\label{prex}\\
&&n_i=\phi_i\hsp\exp\left[-\psi\hspm (2\hspm n_B\hspm v)-
n_M^{\rm id}\hspm v\right],~~~i\hspm\in B,\ov{B},\label{denb1}\\
&&n_i=\phi_i\hsp (1-2\hspm n_B\hspm v)\,,\hspace*{3.1cm}i\hspm\in M,\label{denm1}
\end{eqnarray}
where $n_M^{\rm id}=\sum\limits_{i\hspm\in M}\phi_i$. Note that the meson
component of pressure, given by the last term in the r.h.s.~of~(\ref{prex}),
is the same as in the ideal gas. However, the total pressure
and partial densities are reduced due to the interaction of (anti)baryons with mesons.
Equa\-tions~\mbox{(\ref{denb1}) and (\ref{denm1})} can be easily solved with respect to~$n_i$.
Indeed, taking {a} sum of both sides of~\re{denb1} over all $i\hspm\in B$, one gets
a single transcendental equation for~$n_B$
\bel{denb2}
n_B=n_B^{\rm id}\hsp\exp\left[-\psi\hspm (2\hspm n_B\hspm v)-n_M^{\rm id}\hspm v\right],
\ee
where $n_B^{\rm id}=\sum\limits_{i\hspm\in B}\phi_i$. One can see that
solving~\re{denb2} is sufficient for calculating pressure and partial densities of all hadrons.

\begin{figure*}[bht!]
\centerline{\includegraphics[trim=0 7.5cm 0 8.5cm, clip,
width=0.65\textwidth]{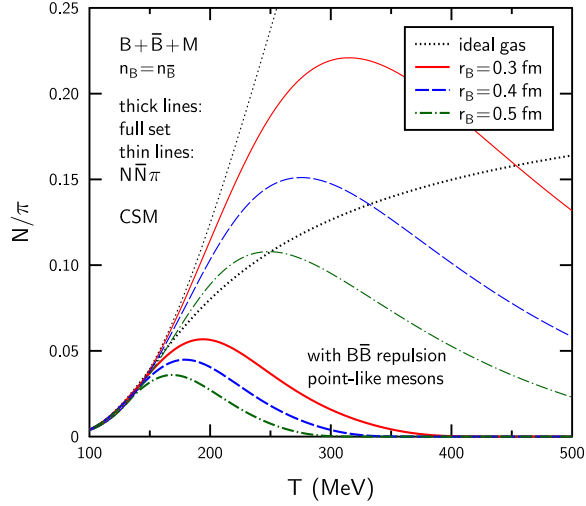}}
\caption[]{(Color online)
The $N/\pi$ ratio as a function of temperature in $B\ov{B}M$ matter
with point-like mesons. Thick lines show the CSM results for the full set of hadrons
with inclusion of feeding from strong decays of resonances (the CI-model).
Thin lines are calculated for the $N\ov{N}\pi$ matter.
The dotted lines represents the ideal gas limiting case.
}\label{f5:nrpi-c1-red}
\end{figure*}
The resulting $N/\pi$ ratios are presented in Fig.~\ref{f5:nrpi-c1-red}. The
$N/\pi$ ratio in the full hadron gas calculation is much smaller than
in the $N\ov{N}\pi$ matter. This follows from the fact that the
number of additional pions due to decays of meson and baryon resonances is much larger
than the corresponding number of nucleons produced in these decays. Evidently, such effects exist already
in the ideal gas limit,~$r_B\to 0$ (compare the dotted lines in Fig.~\ref{f5:nrpi-c1-red}).
As compared to the $N\ov{N}\pi$ matter, the $N/\pi$ ratio in the full calculation is
a much narrower function of $T$ and its maximum is shifted to lower
tempera\-tures~$T\lesssim 200~\textrm{MeV}$.

\begin{figure*}[bht!]
\centerline{\includegraphics[trim=0 7.5cm 0 8.5cm, clip, width=0.65\textwidth]{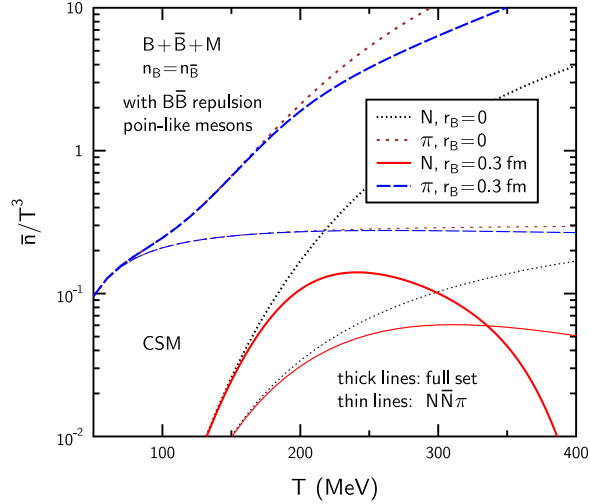}}
\caption[]{(Color online)
Scaled densities $\ov{n}_N/T^3$ and $\ov{n}_{\ds\pi}/T^3$ in $B\ov{B}M$
matter with point-like mesons. Thick lines show the CSM results for the full set of hadrons
(the CI--model). Thin lines are calculated for the $N\ov{N}\pi$ matter.
All calculations correspond to hard-core radius $r_B=0.3~\textrm{fm}$.
The dotted lines represents the ideal gas limiting case.
}\label{f6:nt3-c1}
\end{figure*}
The temperature dependence of scaled hadron densities $\ov{n}_N$ and $\ov{n}_{\ds\pi}$
(with inclusion of feeding from resonance decays) is
given in Fig.~\ref{f6:nt3-c1}. One can see that for all tempe\-ra\-tures~\mbox{$\ov{n}_N\ll\ov{n}_{\ds\pi}$}.
At $T\gtrsim 100~\textrm{MeV}$, the scaled den\-sity~$\ov{n}_{\ds\pi}/T^3$ depends only weakly
on the temperature for the reduced particle set, but it strongly increases with $T$ in the full
hadron gas calculation. Such a behavior is due to significant excitation of heavy mesons at large temperatures.

\subsection{Hadronic matter without baryon-antibaryon repulsion}

\begin{figure*}[bht!]
\centerline{\includegraphics[trim=0 7.5cm 0 8.5cm, clip, width=0.65\textwidth]{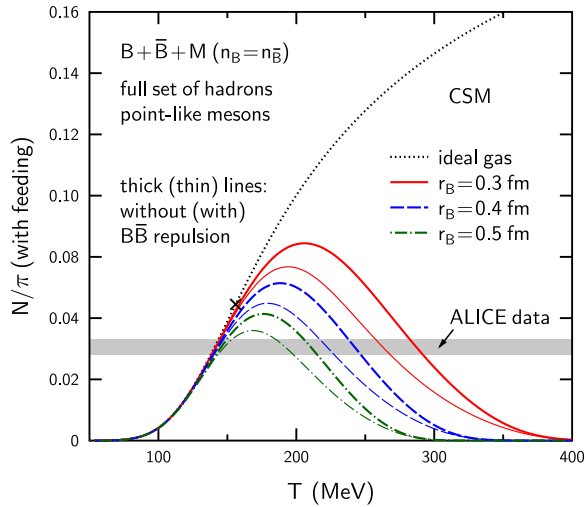}}
\caption[]{(Color online)
{The} $N/\pi$ ratio as a function of temperature in the $B\ov{B}\hspm M$ matter
(the full set of hadrons) with point-like mesons. Thick (thin) lines show the CSM
results within the CII~(CI) model. The dotted line represents the ideal gas limiting case.
Shading shows experimental bounds for the $N/\pi$ ratio obtained~\cite{Abe12} for 0-5\%
central Pb+Pb collisions at $\sqrt{s_{NN}}=2.76~\textrm{TeV}$. The cross represents
the $N/\pi$ value in the ideal gas at $T=156~\textrm{MeV}$.
}\label{f7:nrpi-c1-c2}
\end{figure*}
Let us consider now matter without  $B\ov{B}$ interactions: we refer
to this approach as the CII model. As explained in Appendix~C, in this case we get the
same equations as above, but with the replacement \mbox{$2\hsp n_B\to n_B$} in the arguments
of the functions $Z$ and~$\psi$ in~Eqs.~(\ref{prex})--(\ref{denb1}), (\ref{denb2}). Using the same
arguments as in Sec.~\ref{nnpw} one can show that the $N/\pi$ values should increase in
the CII model as compared to the CI calculation. The numerical results presented
in~Fig.~\ref{f7:nrpi-c1-c2} confirm this conclusion. The shaded region in this figure
shows the ALICE constraint~\cite{Abe12} for central Pb+Pb collisions at the LHC energy.
In fact, this Collaboration gives experimental bounds for $(p+\ov{p})/(\pi^++\pi^-)$. To get
corresponding values of $N/\pi$, we introduce the additional factor $2/3$.

Up to now there are no robust estimates of the hard-core radius $r_B$. It is natural to
assume that $r_B$ is of the order of the nucleon quark-core radius $r_q$\hspm\footnote
{
We assume that all baryons have approximately same sizes of quark cores. It is argued in
Ref.~\cite{Neu75} that repulsion between baryons at small distances $r<2\hsp r_q$ appears
due to the Pauli principle which prevents identical fermions (quarks) to overlap in the phase space.
}.
The chiral bag calculations~\mbox{\cite{Teg83,Myh88}} give $r_q\simeq 0.5-0.6~\textrm{fm}$.
According to our analysis, the $N/\pi$ values calculated in the CII model meet the
experimental constraint for~$r_B\lesssim 0.68~\textrm{fm}$\hsp\footnote
{
At larger $r_B$ the calculated $N/\pi$ values are below the shaded region in Fig.~\ref{f7:nrpi-c1-c2}.
The corresponding condition for the CI--case is $r_B\lesssim 0.62~\textrm{fm}$.
}.
 For example, at~$r_B=0.5~\textrm{fm}$ the CII-curve in Fig.~\ref{f7:nrpi-c1-c2} crosses the
 ''experimental'' strip  at two temperature intervals: $T=147\pm 4~\textrm{MeV}$ and
 $T=209\pm 6~\textrm{MeV}$. Note that in the ideal gas limit $r_B\to 0$ only one (low temperature)
 interval remains.
\begin{figure*}[bt!]
\centerline{\includegraphics[trim=0 7.5cm 0 8.5cm, clip, width=0.65\textwidth]{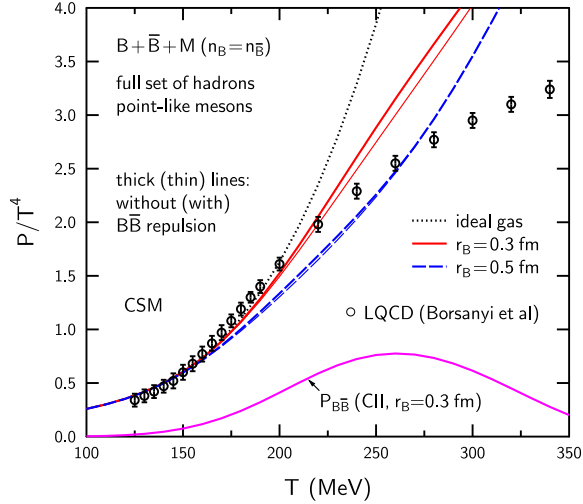}}
\caption[]{(Color online)
Scaled pressure as a function of temperature in the $B\ov{B}\hspm M$ matter
(the full set of hadrons) with point-like mesons. Thick (thin) lines show the CSM results
within the~CII~(CI) model. The dotted line represents the ideal gas limiting case.
The lower solid curve gives the (anti)baryon part of pressure calculated in the CII model
with $r_B=0.3~\textrm{fm}$. Open dots shows results of lattice QCD calculation~\cite{Bor14}.
}\label{f8:pt4-c1-c2}
\end{figure*}

The $N/\pi$ ratio predicted by the ideal gas thermal model~\cite{Bra16} for the same reactions
corresponds to the point $T=156~\textrm{MeV}$ at the ideal gas curve. This point is marked
by a cross in Fig.~\ref{f7:nrpi-c1-c2}. One can see that the thermal model fit overestimates
noticeably the (anti)nucleon-to-pion ratio observed at LHC. Attempts to resolve this discrepancy
by introducing the annihilation and regeneration of $B\ov{B}$ pairs at late stages of
the reaction have been made in Refs.~\cite{Sat13,Pan14}.

In Fig.~\ref{f8:pt4-c1-c2} we compare the results of pressure calculations within the
CI and CII models with the lattice QCD data. One can see these two models predict rather
similar results. The low sensitivity of pressure to the omission of $B\ov{B}$ repulsion is
explained by a relatively small contribution of (anti)baryons (the first term of~\re{prex}) as
compared to mesons.

\subsection{Hadronic matter without (anti)baryon-meson repulsion}

Up to now we described the $MB$ and $M\ov{B}$ interactions
in the excluded volume scenario: it was assumed that point-like mesons do not penetrate into the
volume occupied by hard-cores of (anti)baryons. In this approximation interactions of (anti)baryons
with mesons lead to additional shifts of chemical potentials given by the last
terms in Eqs.~\mbox{(\ref{mub1}) and (\ref{mum1})}. One should have in mind, that such a purely classical
picture is especially questionable for pion interactions. Indeed, at
$T\gtrsim m_{\ds\pi}$ the thermal wave length of pions is of the order of \mbox{$T^{-1}$}.
At~$T\lesssim 400~\textrm{MeV}$ this length is at least comparable with typical hardcore ra\hspm dii of baryons.

\begin{figure*}[bht!]
\centerline{\includegraphics[trim=0 7.5cm 0 8.5cm, clip, width=0.65\textwidth]{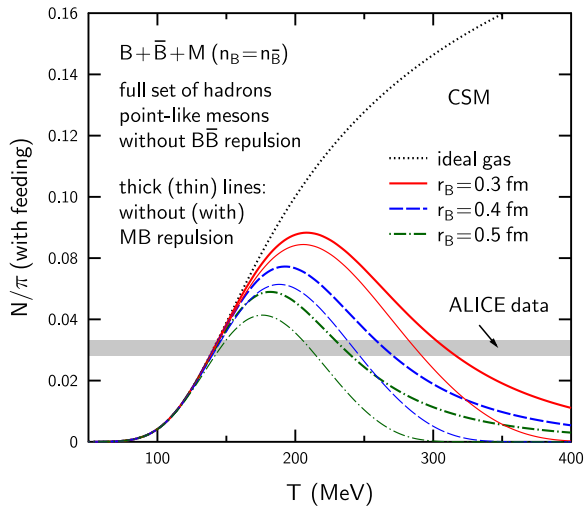}}
\caption[]{(Color online)
The ratio $N/\pi$  as a function of temperature in the $B\ov{B}\hspm M$ matter
(the full set of hadrons) with point-like mesons. Thick (thin) lines show the CSM results within
the~CIII~(CII) model. The dotted line represents the ideal gas limiting case. Shading shows experimental
bounds for the $N/\pi$ ratio obtained~\cite{Abe12} for 0-5\%
central Pb+Pb collisions at $\sqrt{s_{NN}}=2.76~\textrm{TeV}$.
}\label{f9:nrpi-c3-c2}
\end{figure*}
Quantum calculations of of the second viral coefficients have been
performed earlier for purely pion~\cite{Kos01} and $\pi\hspm N\hspm K$~\cite{Ven92} matter. These coefficients
were expressed via the momentum integrals of phenomenological phase shifts
of binary hadronic scattering. It was shown that the repulsive and attractive contributions
nearly cancel for pion interactions. Qualitatively, such interactions may be described
by the addition of meson and baryon resonances\hsp\footnote
{
Note that this conclusion is obtained only for second--order viral coefficients
and therefore, it may be not valid for high-order terms.
}.
Interactions of heavier mesons are, presumably, less modified due to quantum effects. Note that the role
of Pauli suppression for short-range interactions of mesons with (anti)baryons should be less significant
as compared to $BB$ and $\ov{B}\ov{B}$ pairs.

\begin{figure*}[bht!]
\centerline{\includegraphics[trim=0 7.5cm 0 8.5cm, clip, width=0.65\textwidth]{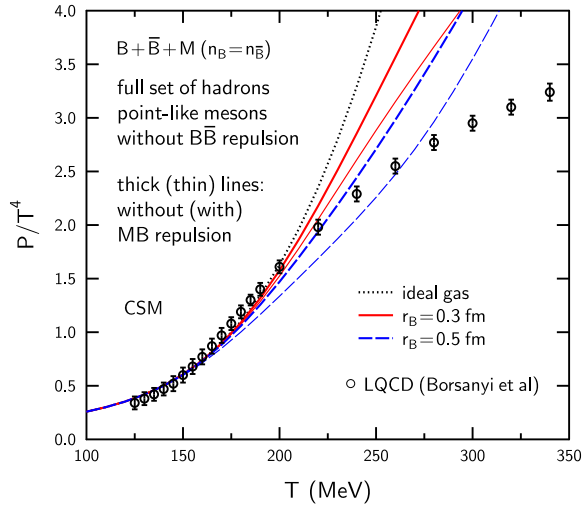}}
\caption[]{(Color online)
Scaled pressure as a function of temperature in the $B\ov{B}\hspm M$ matter
(the full set of hadrons) with point-like mesons. Thick (thin) lines show the CSM results within
the~CIII~(CII) model. The dotted line represents the ideal gas limiting case.
Open dots shows results of lattice QCD calculation~\cite{Bor14}.
}\label{f10:pt4-c3-c2}
\end{figure*}
This discussion shows that one can expect reduced $MB$ and $M\ov{B}$ short-range
interactions as compared to a simple classical scenario. To estimate the role of such interactions, we modify
the~CII calculation, by omitting additionally the $MB$ and $M\ov{B}$ short-range repulsion for all mesons.
In this new scenario (the CIII model) mesons can be considered as an ideal gas,
with partial densities $n_i=\phi_i~(i\hspm\in M)$. Omitting denominators in
Eqs.~(\ref{pec1})--(\ref{mub1}) and applying the procedure similar to those used in
Sec.~\ref{wbbr} we get the relations
\begin{eqnarray}
&&\frac{P}{T}=2\hspm n_B\hspm Z\hspm (n_{\hspm B}\hspm v)+n_M^{\rm id}\,,\label{prex3}\\
&&n_i=\phi_i\hspm\exp\left[-\psi\hspm (n_B\hspm v)\hspm\right],~~~i\hspm\in B,\ov{B},\label{denb3}
\end{eqnarray}
where $n_B$ is determined by solving the equation
$n_B=n_B^{\rm id}\hsp\exp\left[-\psi\hspm (n_B\hspm v)\hspm\right]$. It is clear that the
hadron densities and pressure increase in the new scenario as compared to the CII
model.

Comparison of CII and CIII calculations is given in \mbox{Figs.~\ref{f9:nrpi-c3-c2}
and~\ref{f10:pt4-c3-c2}}.~It is seen that the~$N/\pi$ ratio
increases and becomes a broader function of temperature in the CIII model. These effects
are more significant at larger $r_B$. According to Fig.~\ref{f10:pt4-c3-c2}, the pressure
is closer to the ideal gas in the new calculation, especially at large temperatures.
It is important that at~\mbox{$r_B\sim 0.5~\textrm{fm}$} the lattice QCD data are satisfactory reproduced in the CIII
model at~\mbox{$T\lesssim 250~\textrm{GeV}$}.

\section{Fitting the ALICE hadron yields in Pb+Pb collisions}

In this section we perform the fit to the midrapidity yields of hadrons $\pi^\pm,
K^\pm, K_S^0, p, \ov{p}$, $\Lambda, \Xi^\pm, \Omega^\pm$, and $\phi$ measured by the ALICE
collaboration in the $0-5\%$
central Pb+Pb collisions at $\sqrt{s_{NN}}=2.76$~TeV \cite{ALICE}\hspm\footnote
{
Note that the centrality binning for $\Xi$ and $\Omega$ hyperons is different from other hadrons
in the ALICE experiments. Thus, we take the midrapidity yields of $\Xi$ and $\Omega$ in the $0-5$\% centrality
class from Ref.~\cite{Bec14}, where they were obtained using the interpolation procedure.
}.
We perform the fits for the ideal hadron gas as well as for the
CI, CII, and CIII models, introduced in Sec.~\ref{shrg}. In these calculations
nonzero mass widths of resonances are taken into account (for details, see Ref.~\cite{Vov16}).
The baryon compressibility factor $Z$ is chosen in the excluded volume form (\ref{zem}), and all
mesons are considered as point-like particles.

\begin{figure*}[bht!]
\centerline{\includegraphics[width=0.9\textwidth]{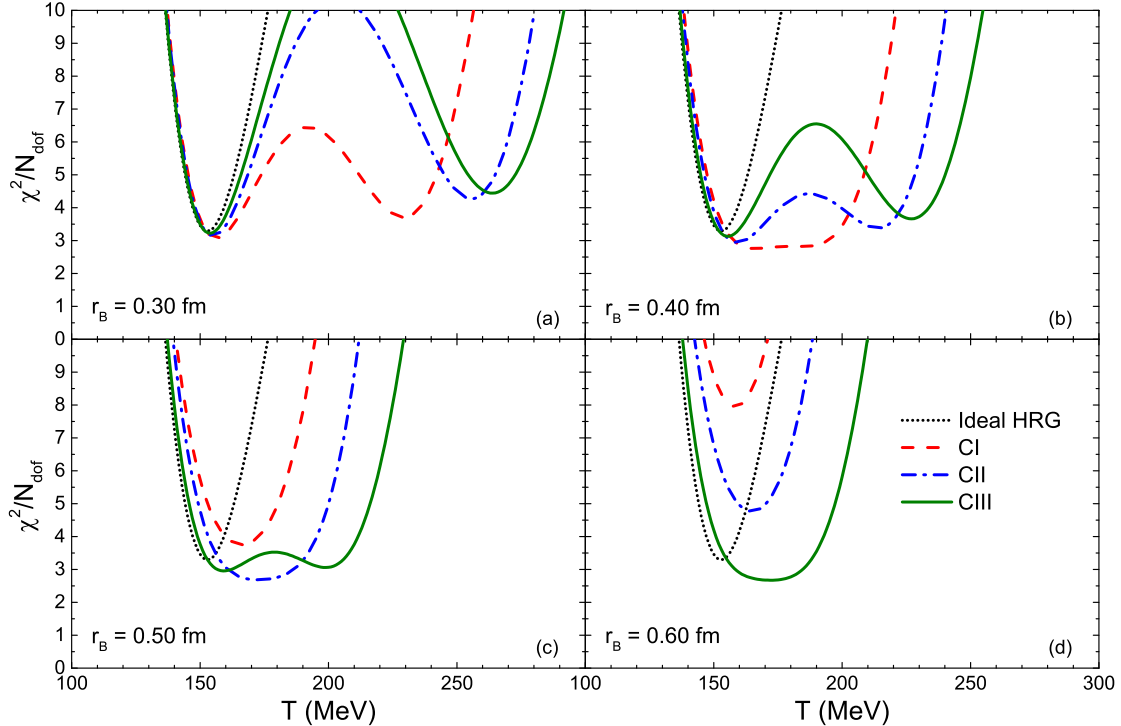}}
\caption[]{(Color online)
{The temperature dependence of $\chi^2/N_{do\hspace*{-1.28pt}f}$ for fitting the ALICE data on
hadron yields in 0-5\%
central Pb+Pb collisions at $\sqrt{s_{NN}}=2.76~\textrm{TeV}$. The dashed, dash-dotted,
and solid curves correspond, respectively, to the models CI, CII, and CIII\hspm .
The dotted lines are calculated in the ideal gas limit $r_B=0$\hspm .}
}\label{f11:chi2-c1-c3}
\end{figure*}

At the LHC energies the asymmetry in the production of particles and antiparticles
becomes negligible. This implies nearly zero values of all chemical potentials in a~statistical system.
The hadron yield ratios would then be determined by a single free parameter -- the
so called chemical freeze-out
temperature $T$. The statistical quality of such a fit is defined by the value of $\chi^2/N_{\rm dof}$.
%The systematic model uncertainties of this fit procedure are sizeable
Figure \ref{f11:chi2-c1-c3} shows the temperature dependence of
$\chi^2/N_{\rm dof}$ for the ideal gas (dotted lines), as well as for the models CI, CII, and CIII.
The results with hard-core baryon
radii $r_B=0.3, 0.4, 0.5$, and $0.6~\textrm{fm}$ are presented in the panels (a), (b), (c), and (d),
respectively. In all models, at each temperature, the only remaining free parameter is the system volume.
In fitting the midrapidity hadron yields, this parameter is fixed
at each $T$ to minimize the $\chi^2$ values of the fit. The best fit for each model
corresponds to the minimum $\chi^2/N_{\rm dof}$ value.

Two important features of the results shown in Fig.~\ref{f11:chi2-c1-c3}
should be pointed out.
First, introducing non-zero hard-core radii for baryons increases  the chemical freeze-out
temperature (i.e. shifts the position of the $\chi^2/N_{\rm dof}$ minimum to the right) and
improves the fit's quality (reduces the minimum  value of $\chi^2/N_{\rm dof}$). Second,
the structure of the $\chi^2/N_{\rm dof}$ curves as a function of~$T$ in the models
CI, CII, and CIII is very different as compared to the ideal gas case. The corresponding curves are
noticeably wider, and in many cases two minima appear.
Therefore, extracting the chemical
freeze-out temperature from the hadron yield data is a~rather delicate procedure: it strongly
depends on details of the model for the short-range repulsion
and on the choice of hard-core radii of baryons. One can see a significant
difference between the CI and CII results. This shows that
improved excluded volume models
should take into account the difference between the $BB$ and $B\ov{B}$ interactions.
Introducing non-zero hard-core radii for mesons $r_M< r_B$ within the NDEM can improve
both the quality of fitting the observed hadron yields and the agreement with the
lattice~QCD data at~$T\sim 200$~MeV.

\section{Conclusions and outlook}

Modeling the short-range interactions in the hadronic gas
remains an open problem. For a simple system of finite-size (anti)baryons
and point-like mesons, here several different formulations (e.g., CI, CII, and CIII) are
analyzed.
%We show that
Omission of meson--baryon and antibaryon-baryon interactions strongly
influences the pressure and hadronic densities.

Similar to Ref.~\cite{Vov16} it is found that commonly used diagonal
eigenvolume models are not realistic for quantitative studies, especially
in the situation when hard-core radii of mesons are significantly smaller
than those for (anti)baryons.

The nucleon-to-pion ratio in hadronic gas is a non-monotonic
function of temperature. This increases uncertainties
in attempts to extract  the freeze-out temperature by
fitting the hadronic ratios observed in heavy-ion collisions. This also explains
the appearance of the second, high-temperature minimum of $\chi^2$ distribution
obtained from the thermal fit of~ALICE data in Ref.~\cite{Vov15a}.

We show that more refined formulations
of the eigenvolume model (e.g. the Carnahan-Starling one) give similar results
for baryon-symmetric matter as compared to the simple van der Waals approach.

Simultaneous description of the observed hadron ratios and lattice QCD data
for the pressure, energy density, as well as fluctuations of conserved charges
may help in studying the role of short-range interactions in hadronic matter. Of course,
the transition region between the fully confined hadronic phase and the deconfined matter
should be taken into account at high enough temperatures.
To take these effects into account, one can use
the approach suggested in Ref.~\cite{Alb14}. In this scheme a~crossover EoS has been
suggested which interpolates the hadron gas phase with excluded volume corrections and
the deconfined states in a perturbative QCD model.

\begin{acknowledgments}
The authors thank K. A. Bugaev, I. N. Mishustin, and P. M. Lo
for useful discussions. A~partial support from the grant NSH--932.2014.2
of the Russian Ministry of Education and Science
is  acknowledged by L.M.S.
The work of M.I.G. is supported by the
Goal-Oriented Program of  the National Academy of Sciences of Ukraine and the
European Organization for Nuclear Research (CERN), Grant CO-1-3-2016,
and by the Program of Fundamental Research of the Department of Physics and
Astronomy of National Academy of Sciences of~Ukraine.

\end{acknowledgments}

\appendix
\section{}
\label{app-A}

Since the particle numbers are not fixed in the equilibrium hadronic matter,
it is natural to describe its properties in the grand canonical variables. On the other hand,
the hard-sphere interactions are easier introduced in the CE. Below we apply a rather general algorithm
of transition to the GC\hspm E suggested in Refs.~\cite{Sat15,Ald55}. An equivalent procedure
has been used in~\cite{Vov15b,Red16}.

Let us assume that one knows pressure $P$ in the CE, i.e. as a function of temperature and partial
densities~$n_i$. Then one can easily calculate the free energy density $f=F/V=f\hsp (T,n_1,n_2,\ldots)$
for arbitrary multiparticle interactions (see below). This quantity is a genuine thermodynamical potential
in the CE\hspm . One can write the thermodynamic relation~\cite{Lan75}:
\bel{trfp}
f=\sum\limits_i\mu_i\hsp n_i-P\,,
\ee
where $\mu_i=(\partial f/\partial n_i)_T$ is the $i$\hspm th particle chemical potential and the sum
goes over all particle species $i$\hspm . In the limit of ideal Boltzmann gas one has the following
relations for thermodynamic functions in the CE (see~\re{idgp})
\bel{idg1}
P^{\hsp\rm id}=T\sum\limits_i n_i,~~~~\mu^{\rm id}_i=T\ln{\frac{n_i}{\phi_i}},~~~~f^{\rm id}=
T\sum\limits_i n_i\left(\ln{\frac{n_i}{\phi_i}}-1\right).
\ee
Here the last equality follows from~\re{trfp}.

Particle interactions give rise to nonzero shifts of the free energy density,
$\Delta f=f-f^{\rm id}$, and pressure, $\Delta P=P-P^{\hsp\rm id}$, with respect
to the ideal gas. One can use the relation $dF=Pd\hspm V$ for the free
energy change in the isothermal process. Taking the integral over the system volume $V$ (at fixed
particle multiplicities $N_i=n_i\hspm V$), we get the equation~\cite{Sat15}
\bel{fesh}
\Delta\hspm f\hsp (T,n_1,n_2,\ldots)=\int\limits_0^1\frac{d\alpha}
{\alpha^2}\,\Delta\hspm P\hsp (T,\alpha\hsp n_1,\alpha\hsp n_2,\ldots)\,.
\ee
Finally, the free energy density is found from $f=f^{\rm id}+\Delta f$.

Let us consider first the one-component matter with hard-sphere interactions. In accordance
with~\re{pbm} one has \mbox{$\Delta P\hsp (T,n)=nT\left[Z(n\hspm v)-1\hspm\right]$}, where $v$ is the
hard-core volume of a~single particle. Substituting this expression into~(\ref{fesh}) one obtains
\bel{fesh1}
%\textcolor
\Delta\hspm f\hsp (T,n)=nT\int\limits_0^{n\hspm v}\frac{d\eta}{\eta}\,\left[Z(\eta)-1\right].
\ee

Taking further the density derivative, one gets the following expression for the shift of chemical potential:
\bel{cpsh}
\Delta\hsp\mu=\left(\frac{\partial\Delta f}{\partial\hsp n}\right)_T=
T\hspm\psi\hspm (n\hspm v),
\ee
where $\psi\hspm (\eta)$ is defined in~\re{psi}. One can see, that $T\psi\hsp (\eta)$ gives the chemical
potential shift for a one-component gas with the packing
fraction $\eta=n\hspm v$. Formally, this shift can be described by introducing
the repulsive mean-field potential $U\hspm (T,n)=T\psi\hspm (n\hspm v)$~\cite{Anc15}.
Adding the ideal gas chemical potential (see~\re{idg1}) gives~\cite{Sat15}
\bel{cpoc}
\mu=T\left[\ln{\frac{n}{\phi\hsp (T)}}+\psi\hspm (n\hspm v)\hspm\right] .
\ee
This formula can be written in the form $n=n^{\rm id}(T,\mu-\Delta\hsp\mu)$, which is in turn
equivalent to~\re{ngce} of the main text. Finally, pressure $P=P\hsp (T,\mu)$ in the GC\hspm E is
obtained by substituting in (\ref{pbm}) the solution of~\re{cpoc} with respect to $n$\hspm .

By using \re{psiem} one can show that \re{cpoc} leads to the following equation for pressure
in the EVM:
\bel{prev2}
P=P^{\rm id}\left(T,\mu-b\hspm P\right)\,,
\ee
where $b=4\hspm v$ and $P^{\,\rm id}(T,\mu)=T\phi\hsp (T)\exp{(\mu/T)}$.
Note that this model becomes inaccurate at densities $n\gtrsim 0.2/v$.

Let us now consider the GC\hspm E formulation of the DEM introduced in
Sec.~\ref{sdem}\hspm . Substuting $\Delta P=T\sum_in_i\left[(1-\sum_jb_jn_j)^{-1}-1\right]$
into \re{fesh1} gives the relations
\begin{eqnarray}
&&\Delta f=T\sum\limits_i n_i\ln{\frac{\xi_i}{n_i}}\,,\label{dfdem}\\
&&\Delta\mu_{\hspm i}=\left(\frac{\partial\Delta f}
{\partial\hsp n_i}\right)_T=T\ln{\frac{\xi_i}{n_i}}+
b_{\hspm i}\hspm P\,.\label{dmdem}
\end{eqnarray}
Adding the ideal gas chemical potential from~\re{idg1} gives \re{cpdem} of the main text.

 A similar transition algorithm can be developed in the NDEM. Starting from~\re{pnds} and
 calculating the integral in~\re{fesh}, one obtains the relations (\ref{cpnds}) and (\ref{xinds}).

\section{}
\label{app-B}

Let us consider a two-component matter composed of particles with hard-core
radii~$r_1,r_2$ and assume that $r_2=0$. The EoS for such matter in the canonical
variab\-les~$T,n_1,n_2$ is given by~\re{pbm1}. We use a general form of the compressibility
factor for the first component $Z=Z\hspm (\eta_1)$, where $\eta_1=n_1v_1$ and $v_1=4\pi r^3_1/3$.
The results in the EVM and CSM are obtained after sub\-sti\-tu\-ting~\mbox{$Z=Z_{\rm EV}$} and
\mbox{$Z=Z_{\rm CS}$}, respectively.

Using Eqs.~(\ref{pbm1}) and (\ref{fesh}) one gets the relation for the free energy shift
\bel{fesh2}
\Delta f=Tn_1\int\limits_0^{\eta_1}\frac{d\eta}{\eta}\,\left[Z(\eta)-1\right]
-Tn_2\ln{(1-\eta_1)}\,.
\ee
This equation leads to the following formulae for shifts of chemical potentials:
\begin{eqnarray}
&&\Delta\hsp\mu_1=\left(\frac{\partial\Delta f}{\partial\hsp n_1}\right)_T
=T\left[\psi\hspm (\eta_1)+\frac{n_2\hsp v_1}{1-\eta_1}\right],\label{dcpt1}\\
&&\Delta\hsp\mu_2=\left(\frac{\partial\Delta f}{\partial\hsp n_2}\right)_T
=-T\ln{(1-\eta_1)}\,.\label{dcpt2}
\end{eqnarray}
Adding the ideal gas chemical potentials $\mu_i^{\rm id}=T\ln{(n_i/\phi_i)}$ gives
Eqs.~(\ref{cpot1}) and (\ref{cpot2}) of the main text.

\section{}
\label{app-C}

Let us consider the hadronic matter with hard-sphere interactions and assume that
all (anti)baryons has the same hard-core radii $r_B$, but mesons are point-like. We
start from the~CI~model which does not distinguish the $BB$ and $B\ov{B}$ interactions.
In this case \mbox{using}~\re{pec1} leads to the relation
\bel{dpec1}
\Delta\hspm P=P-(n_T+n_M)\hsp T=T\hspm n_T\left[Z\hspm (n_T\hspm v)-1\hspm\right]
+Tn_M\left[(1-n_T\hspm v)^{-1}-1\right].
\ee
Calculating the integral in~(\ref{fesh}) and using (\ref{psi}) give the following
expression for the free energy density:
\begin{eqnarray}
f=&&f^{\hsp\rm id}+\Delta f=T\hspace*{-2mm}\sum\limits_{i\hspm\in B,\ov{B},M}
n_i\left[\ln{\frac{n_i}{\phi_i}}-1\right]\nonumber\\
&&+Tn_T\left[\hsp\psi\hspm (n_Tv)-Z\hspm (n_Tv)+1\hsp\right]
-Tn_M\ln{(1-n_Tv)}.\label{fen2}
\end{eqnarray}
The first terms in this equation give the free energy density
of the ideal hadronic gas. Taking the derivative of (\ref{fen2}) with
respect to $n_i$ one obtains \mbox{Eqs.~(\ref{mub1}) and (\ref{mum1})} for chemical potentials.

Analogous formulae for the CII model (without the baryon-antibaryon repulsion)
are given by Eqs.~(\ref{dpec1}) and (\ref{fen2}), with the replacements
\begin{eqnarray}
&&n_T\hspm Z\hspm (n_T\hspm v) \to n_B\hspm Z\hspm (n_B\hspm v)+
n_{\ov{B}}\hspm Z\hspm (n_{\ov{B}}\hspm v),\label{c1z}\\
&&n_T\hspm \psi\hspm (n_T\hspm v) \to n_B\hspm \psi\hspm (n_B\hspm v)+
n_{\ov{B}}\hspm \psi\hspm (n_{\ov{B}}\hspm v).\label{c1p}
\end{eqnarray}
In this case one obtains the same equations (\ref{prex})--(\ref{denb2}), but
with the replacement $2\hspm n_B\to n_B$ in the arguments of functions $Z$ and $\psi$.

\end{document}